%% file: paper.tex
\documentclass[sigconf]{acmart}

\usepackage{tabularx}
\usepackage[utf8]{inputenc}
\usepackage{multirow}
\usepackage{soul}
\usepackage{xcolor}
\usepackage{xspace}
\usepackage{url}
\usepackage{graphicx}
\usepackage{listings,multicol}
\usepackage[caption=false]{subfig}  
\usepackage[export]{adjustbox}
\usepackage{amsmath}
\usepackage{textcomp}
\usepackage{listings}
\usepackage{cleveref}
\usepackage{algorithm}
\usepackage{enumitem}
\usepackage{tcolorbox}
\usepackage{booktabs}
\usepackage{siunitx}
\usepackage{pifont}
\usepackage[noend]{algpseudocode}
\usepackage{balance}

\PassOptionsToPackage{svgnames}{xcolor}

\include{writing-support}

\title[A Large-Scale Comparison of Python Code in Jupyter Notebooks and Scripts]{A Large-Scale Comparison of Python Code \\ in Jupyter Notebooks and Scripts}

\author{Konstantin Grotov}
\affiliation{
  \institution{\textit{JetBrains Research}}
  \institution{\textit{ITMO University}}
}
\email{konstantin.grotov@gmail.com}

\author{Sergey Titov}
\affiliation{
  \institution{\textit{JetBrains Research}}
}
\email{sergey.titov@jetbrains.com}

\author{Vladimir Sotnikov}
\affiliation{
  \institution{\textit{JetBrains Research}}
}
\email{vladimir.sotnikov@jetbrains.com}

\author{Yaroslav Golubev}
\affiliation{
  \institution{\textit{JetBrains Research}}
}
\email{yaroslav.golubev@jetbrains.com}

\author{Timofey Bryksin}
\affiliation{
  \institution{\textit{JetBrains Research}}
}
\email{timofey.bryksin@jetbrains.com}

\begin{document}

\begin{abstract}
    In recent years, Jupyter notebooks have grown in popularity in several domains of software engineering, such as data science, machine learning, and computer science education. Their popularity has to do with their rich features for presenting and visualizing data, however, recent studies show that notebooks also share a lot of drawbacks: high number of code clones, low reproducibility, etc.
    
    In this work, we carry out a comparison between Python code written in Jupyter Notebooks and in traditional Python scripts. We compare the code from two perspectives: \textit{structural} and \textit{stylistic}. In the first part of the analysis, we report the difference in the number of lines, the usage of functions, as well as various complexity metrics. In the second part, we show the difference in the number of stylistic issues and provide an extensive overview of the 15 most frequent stylistic issues in the studied mediums. Overall, we demonstrate that notebooks are characterized by the lower code complexity, however, their code could be perceived as more entangled than in the scripts. As for the style, notebooks tend to have 1.4 times more stylistic issues, but at the same time, some of them are caused by specific coding practices in notebooks and should be considered as false positives. With this research, we want to pave the way to studying specific problems of notebooks that should be addressed by the development of notebook-specific tools, and provide various insights that can be useful in this regard.
\end{abstract}

\maketitle

\input{sections/01-introduction}
\input{sections/02-background}
\input{sections/03-methodology}
\input{sections/04-library}
\input{sections/05-dataset}
\input{sections/06-results}
\input{sections/07-threats}

\input{sections/08-conclusion}

\balance

\bibliographystyle{ACM-Reference-Format}
\bibliography{paper}

\end{document}

%% file: writing-support.tex
\newcommand{\ie}{\emph{i.e.,}\xspace}
\newcommand{\eg}{\emph{e.g.,}\xspace}

\newcommand{\libname}{\textsc{Matroskin}\xspace}

\newcommand{\npavg}{number of arguments per function in code}

\usepackage{xparse}

\NewDocumentCommand{\set}{o m}{%
  \IfNoValueTF{#1}
    {\{#2\}}
    {\{#1 \mid #2\}}%
}

\newcommand{\shortcouplingformula}{
\begin{equation*}
    Coupling_\text{functions} = \sum\limits_{f_i, f_j \in F}\; \sum\limits_{\alpha \in Y_i, \; \beta \in Y_j}\delta_{\alpha\beta}
\end{equation*}
\begin{equation*}
    Coupling_\text{cells} = \sum\limits_{c_i, c_j \in C}\; \sum\limits_{\alpha \in X_i, \; \beta \in X_j}\delta_{\alpha\beta}
\end{equation*}
\begin{equation*}
    \text{where} \; \delta_{\alpha \beta} = 1 \; \text{if } \alpha = \beta \; \text{and } 0 \text{ otherwise.}
\end{equation*}
}

\definecolor{RQ_title}{RGB}{100, 100, 100}
\definecolor{RQ_body}{RGB}{240, 240, 240}
\definecolor{APA_stats}{RGB}{100, 100, 120}

    {\endtcolorbox}
    
\newlist{researchquestions}{enumerate}{1}
\setlist[researchquestions]{label*=\textbf{RQ\arabic*}}

\newcounter{observation}
\newcommand{\observation}[1]{\refstepcounter{observation}
	\begin{center}
		\framebox{
			\begin{minipage}{0.93\columnwidth}
				{} \textit{#1}
			\end{minipage}
		}
	\end{center}
}

\newcommand{\APAstats}[2]{\textcolor{APA_stats}{(M=#1, SD=#2)}}
\newcommand{\APAstatsNoBraces}[2]{\textcolor{APA_stats}{M=#1, SD=#2}}
\newcommand{\APAstatsNoBracesMedian}[3]{\textcolor{APA_stats}{M=#1, SD=#2, Median=#3}}
\newcommand{\APAt}[2]{\textcolor{APA_stats}{t(N)=#1, p $\leq$ #2}}

%% file: sections/01-introduction.tex
\section{Introduction}\label{sec:introduction}

Computational notebooks are a modern implementation of the literate programming paradigm introduced by Knuth in the 1980s~\cite{knuth1984literate}. A distinctive feature of a computational notebook is that the source code of the program can be combined with rich formatted text, pictures, and other media material in order to make the code more comprehensible. One of the most popular types of computational notebooks is Jupyter Notebook\footnote{From here on out, when we say \textit{notebook} in the paper, we mean \textit{Jupyter notebook}.} --- a development environment for creating computational notebooks for Julia, Python, and R programming languages (as well as several others). 

The number of Jupyter notebooks has been growing rapidly over the past few years~\cite{perkel2018jupyter}, they became very popular in the fields of scientific computing, machine learning, data analysis~\cite{perkel2018jupyter}, as well as for creating educational materials~\cite{johnson2020jupyter}. While the media-rich format of notebooks is beneficial for these domains, it also has its downsides. Studies have shown that notebooks have low reproducibility rates~\cite{pimentel2019large}, problems with the execution environment~\cite{wang2021restoring}, high amount of code duplication~\cite{koenzen2020code}, and other drawbacks~\cite{chattopadhyay2020s}. 

However, notebooks are not merely a tool for software development, but a brand new format of code representation. Their unique features, \eg chunking the code into executable cells and different kinds of output, are affecting the coding practices~\cite{rule2018exploration}. Because of these features, we hypothesise that the code written in the notebooks will differ from the code written in traditional scripts.\footnote{In this paper, the term \textit{script} relates to any \texttt{.py} file.} This also leads us to the suggestion that common tools used in software engineering could be misaligned with the notebook code and require adaptation for this novel format.    

To study the specifics of code in computational notebooks, we decided to compare the application of different mediums to the Python code. 
The interpretable nature of Python and the ease of its use make it one of the most popular languages for computational notebooks~\cite{pimentel2019large}. Additionally, Python is widely used in production systems~\cite{python2020survey} and has a large open-source software codebase. In order to check how the medium affects the code, we collected two large datasets --- of Jupyter notebooks and of Python scripts. 

Firstly, we collected all Jupyter notebooks on GitHub as of November 2020, resulting in a total of 9,719,569 \textit{.ipynb} files. After filtering only notebooks with permissive licences and Python as a specified language, we ended up with a dataset of 847,881 notebooks. Before carrying out the main part of our analysis, we evaluated the representativeness of our licensed dataset by comparing a set of structural metrics against the remaining unlicensed notebooks. 
To facilitate further studies in this area, we release this dataset to researchers and practitioners. To collect the dataset of Python scripts, we downloaded 10 thousand most-starred Python projects on GitHub, selected the ones with permissive licenses, and then collected all 465,776 Python scripts within them. 

We compare these two codebases in two dimensions: \textit{structural} and \textit{stylistic}. For the structural analysis, we calculated and compared a set of 15 metrics from traditional software engineering, like cyclomatic complexity and the number of imported functions. One of the challenges here was adapting the metrics to the notebooks, since most of the works in this field targeted traditional scripts and libraries. As a result, we developed a Python library called \libname that facilitates a more convenient way of working with notebooks and allows to calculate their structural metrics.

For the second, stylistic, dimension, we used the Hyperstyle tool~\cite{birillo2022hyperstyle}, which allows running linters from several Python style guides and provides a convenient categorization for the discovered issues. Since the employed style guides were developed for the traditional Python code, this also allows us to check their adequacy when applied to notebooks. Due to the computational intensity of using Hyperstyle, we carried out this analysis on two samples of 100,000 files from each dataset.
We compared the profile of stylistic issues between the notebooks and traditional Python scripts in three categories: (1) \textit{error-proneness} --- violations that could lead to potential bugs, (2) \textit{code style} --- violations of the language style guide, in our particular case --- PEP-8~\cite{pep8}, and (3) \textit{best practices} --- violations of the widely accepted recommendations in Python.

Structurally, we found that code in notebooks differs from its traditional counterpart in several specific aspects. The notebooks demonstrated a distinct pattern in the usage of functions, and the complexity metrics showed that notebooks contain structurally simpler, but more entangled code. 
This could lead to notebooks being less comprehensible. 
As for the stylistic differences, they are more prominent: notebooks have 1.4 times more stylistic errors than scripts. In all three categories, most of the top issues prevailed in the notebooks, however, we show that some of the detected stylistic errors should not be counted as violations in the context of a notebook. For example, one of the most popular stylistic errors is \textit{found statement that has no effect}, which can appear when printing the data or plotting a figure. From the perspective of a linter, these statements have no effect, but from the perspective of a notebook, they represent its core functionality.

Overall, we argue that notebooks significantly differ from scripts in both structural and stylistic ways. This leads us to the main conclusion: notebooks as a medium have their own unique problems that should be addressed by notebook-specific tools. It is important to adapt linters for the notebook functionality, as well as to develop tools that could help manage the excessive entanglement of the notebooks.

The main contributions of our work are the following:

\begin{itemize}
    \item \textbf{Dataset.} We collected the dataset of 847,881 preprocessed, permissively licensed, and representative notebooks for further research. The dataset is available online on Zenodo: \url{https://doi.org/10.5281/zenodo.6383115}.
    \item \textbf{Library.} We developed a Python library called \libname for the preprocessing, storing, and structurally analyzing large datasets of Jupyter notebooks. \libname is available online:  \url{https://github.com/JetBrains-Research/Matroskin}.
    \item \textbf{Analysis.} We provide a comparison between Jupyter notebooks and traditional Python scripts. We discovered that notebooks are structurally simpler and have more stylistic issues, which indicates the need for notebook-specific tools.
\end{itemize}

The rest of the paper is organized as follows. Section~\ref{sec:background} describes the existing works that study notebooks from the structural and the stylistic perspectives. Then, Section~\ref{sec:methodology} describes the overall methodology of our study, Section~\ref{sec:library} presents \libname, the library for processing Jupyter notebooks, and Section~\ref{sec:datasets} explains the pipeline for collecting the datasets. In Section~\ref{sec:findings}, the main findings of our work are listed and discussed. Finally, Section~\ref{sec:threats} describes the threats to the validity of our study, and Section~\ref{sec:conclusion} concludes the paper and discusses possible future work.

%% file: sections/02-background.tex
\section{Background}\label{sec:background}

In the last decade, Jupyter notebooks became quite a popular development environment for several activities in software engineering, such as building machine learning models~\cite{perkel2018jupyter}, creating educational materials~\cite{johnson2020jupyter}, or prototyping various applications~\cite{rule2018exploration}. The most popular language that Jupyter notebooks support is Python~\cite{pimentel2019large}. 

There exist different approaches to studying source code~\cite{caulo2020taxonomy}. We chose to separate our analysis into two logical parts: (1) \textit{structural analysis} --- which includes the raw properties of the code, such as complexity metrics, function counts, etc., and (2) \textit{stylistic analysis} --- which is focused on inefficient coding patterns and styling violations. Let us now overview the existing works in these areas. 

\subsection{Structural Studies on Code}

Structural analysis of code is a well-studied area. During the decades of the static analysis of code, many metrics were engineered for measuring such structural properties as code complexity~\cite{mullanu2020code} or maintainability~\cite{coleman1994using}. The most exhaustive set of metrics was presented in the work of Caulo et al.~\cite{caulo2020taxonomy} --- the authors collected over 300 different metrics for code evaluation. While some of these metrics are not applicable for Python, there are several works that focused specifically on evaluating Python code. 

The most recent work in this field was carried out by Pend et al.~\cite{peng2021empirical}. The authors focus on different language features that are used in Python code, and show that decorators, \texttt{for} loops, and nested classes are among the most used features.  

Structural analysis is a part of one of the most exhaustive studies of Jupyter notebooks carried out by Piementel et al.~\cite{pimentel2021understanding}. In the paper, the authors focus on the reproducibility issue of the notebooks --- as they report in their previous work~\cite{pimentel2019large}, only 20\% of notebooks could be fully executed after cloning, and even fewer of them could reproduce previous results. 
The authors provide an analysis of how people use the literate features of the notebooks, \textit{e.g.,} Markdown cells, and, what is more important for us, they perform a simple analysis of code in the notebooks. 
They carry out an analysis of the import statements, as well as the number of defined classes and functions. 
As a result, they show that only a fraction of notebooks (8.54\%) define their own classes, and that notebooks are mainly designed to orchestrate sets of imported functions. 
In our work, we want to deepen our understanding of these phenomena and investigate more of how the notebook code is designed.

While some tools facilitate the calculation of structural metrics, \eg Radon~\cite{radon} or Beinget~\cite{beniget}, none of them support the format of notebooks, so we had to develop our own solution.  

\subsection{Stylistic Studies on Code}

Searching for code style violations is a very interesting academic task, as well as an important practical one. Low stylistic quality is one of the predictors of the low reproducibility of code~\cite{martin2009clean} and a higher percent of test failings~\cite{glass2002facts}. 

There are a number of production-oriented tools for evaluating code style, from automated fixes in integrated development environments (IDEs) such as IntelliJ IDEA~\cite{idea} or VS Code~\cite{vscode}, up to standalone utilities like \textit{flake8}~\cite{flake8} or its improved counterpart --- \textit{Wemake Python Styleguide}~\cite{wemake}. 

There are numerous works on the topic of code quality~\cite{nundhapana2018enhancing, keuning2017code}, however, there exists a lack of large-scale studies of code style violations in Python. In a recent paper, Simmons et al.~\cite{simmons2020large} analysed the differences in styling violations between production Python code and Data Science projects. The authors showed that Data Science projects are more likely to have the number of local variables be more than recommended, or have more arguments in functions. 

As for the stylistic analysis of Jupyter notebooks, Wang et al.~\cite{wang2020better} carried out such a study on a relatively small sample of 1,947 notebooks. They reported some of the descriptive features in the notebooks, like lines of source code (SLOC) and the number of cells. They also reported how well the notebooks are written from the perspective of the Python stylistic requirements. The authors found that notebooks had significantly more PEP8 violations compared to traditional scripts in the same repository. 

We already mentioned several tools for the stylistic analysis of Python code, like \textit{flake8} and \textit{WeMake}. The main difference between these tools is that they are based on different style guides and therefore provide a different set of reported issues. However, there exist educational tools~\cite{keuning2019teachers, birillo2022hyperstyle} that combine different style guides and provide a common classification of different errors. One such tool is Hyperstyle~\cite{birillo2022hyperstyle}, it combines several style guides and outlines five categories of issues: \textit{Code style}, \textit{Code complexity}, \textit{Error-proneness}, \textit{Best practices}, and \textit{Minor issues}. While the original authors used these categories for scoring student submissions, we find them to be helpful in the empirical analysis of the Jupyter notebooks.

\subsection{Summary}

While there exists research on the differences between various subsets of the Python codebase, there are no large-scale comparison studies between Python code in the form of a notebook and a script. In our work, we strive to achieve a comprehensive description of notebook specifics. Analysing the code's both structural and stylistic perspectives will allow us to make conclusions about the differences in the composition and the quality of code in notebooks, as well as how code inspection tools should be adapted to them. 

\input{sections/03-methodology-table}

%% file: sections/03-methodology-table.tex
\begin{table*}[]
\begin{tabular}{clp{8cm}cc}
\toprule

  \textbf{No.} &\multicolumn{1}{c}{\textbf{Metric name}} &
  \multicolumn{1}{c}{\textbf{Description}} &
  \textbf{Agg.} &
  \textbf{Norm.} \\ 
  \midrule\midrule
  \multicolumn{5}{c}{\textbf{Code writing}}\\\midrule\midrule

  \textbf{1} & SLOC &
  Number of source code lines &
  Sum  &
  No \\\midrule

  \textbf{2} & Comment LOC &
  Number of comment lines in the code &
  Sum  &
  Yes \\\midrule

  \multirow{2}{*}{\textbf{3*}} & \multirow{2}{*}{Extended comment LOC} &
  \textit{Only for notebooks}: Number of comment lines in the code &
  \multirow{2}{*}{Sum} &
  \multirow{2}{*}{Yes} \\
  &&  combined with number of lines in adjacent Markdown cells & & \\\midrule

  \textbf{4} & Blank LOC &
  Number of blank lines in the code &
  Sum  &
  Yes \\
  \midrule\midrule
  \multicolumn{5}{c}{\textbf{Function usage}}\\\midrule\midrule

  \textbf{5-6} & Built-in functions (unique and count) &
  Number of Python's built-in functions (unique and total) &
  Sum  &
  Yes \\\midrule
 
  \textbf{7-8} &User-defined functions (unique and count) &
  Number of user-defined functions (unique and total) &
  Sum  &
  Yes \\\midrule
 
  \textbf{9-10} & API functions (unique and count) &
  Number of directly imported functions (unique and total) &
  Sum  &
  Yes \\\midrule
 
  \textbf{11} & Other functions (count) &
  Number of all the remaining functions (total) &
  Sum  &
  Yes \\
  \midrule\midrule
\multicolumn{5}{c}{\textbf{Complexity}}\\\midrule\midrule
  \multirow{2}{*}{\textbf{12}} & \multirow{2}{*}{Cyclomatic code complexity} &
  Number of linearly independent paths through  &
  Max  &
  No \\
  && the program's source code &&\\\midrule

  \multirow{2}{*}{\textbf{13}} &\multirow{2}{*}{Function coupling} &
  Average number of common elements  &
  Avg  &
  No \\
  &&between functions in the code&&\\\midrule

  \multirow{2}{*}{\textbf{14*}} &\multirow{2}{*}{Cell coupling} &
  \textit{Only for notebooks}:  Average number of common  &
  --- &
  No \\
  &&elements between cells in a notebook&&\\\midrule

  \textbf{15} &NPAVG &
  Average number of arguments per function in the code &
  Avg  &
  No \\\bottomrule
\end{tabular}
\vspace{0.3cm}
\caption{The list of the 15 used metrics, divided into three groups. \textit{Agg.} indicates the type of the aggregation function used: sum, average, or maximum. \textit{Norm.} indicates whether this metric is normalized to the number of source lines of code. The star near the metric's number indicates that this metric was proposed by us.}
\vspace{-0.3cm}
\label{table:metrics-description}

\end{table*}

%% file: sections/03-methodology.tex
\section{Methodology}\label{sec:methodology}

To highlight the specifics of notebooks, we had to select the comparison metrics. We decided to study the difference from two perspectives: \textit{structural} and \textit{stylistic}. Both of these directions are represented by hundreds of metrics, and in this section, we will go through the filtration process for each part of the research, and describe the resulting set of metrics.

\subsection{Structural Metrics} \label{sec:methodology_structural}

Firstly, we analyzed the existing papers about structural metrics in software engineering~\cite{isong2013systematic, radjenovic2013software}, most notably, the work of Caulo et al. that presented the list of 300 such metrics~\cite{caulo2020taxonomy}. Since most of the structural metrics were originally created for analysing object-oriented programming code, it is an additional challenge to apply these metrics to notebooks, where classes are rarely used~\cite{pimentel2021understanding}. Thus, we decided that the used metrics had to fulfill two conditions: (1) they have to be applicable to Python, and (2) they have to be applicable to a single notebook or a single cell. 

We chose 13 metrics form the work of Caulo et al. that we believe are meaningful for notebooks and scripts, and that will provide us with information about the structure of the code in these mediums. The full list of metrics is presented in Table~\ref{table:metrics-description}. 
Among the suggested metrics, we found a set of metrics dedicated to counting various types of functions. To analyze this aspect of the notebooks and scripts, we had to detect functions within them. Firstly, we divided all functions into the following groups: \textit{built-in} functions, \textit{user-defined} functions, and \textit{API} functions. Built-in Python functions were found via the comparison with their full list from the Python documentation. User-defined functions were counted as all functions that were directly defined in this particular file. Finally, we counted API functions as functions that were directly imported in the file, or that are the methods of imported modules.
In order to find the functions, we used the abstract syntax tree (AST) representation of code. We detected all the tree nodes that correspond to functions, and collected them, while also checking if each function is from a list of built-ins, is defined in the file, or is imported.
However, in such categorization, not all functions are taken care of. For example, some functions can be imported implicitly (\eg \texttt{import *}). In order not to lose the data about such functions, we calculated them separately as \textit{Other} functions.

All the evaluated metrics were developed for scoring code in the form of traditional scripts, so we additionally introduced two versions of the metrics developed specially for notebooks. Two main features of the notebooks are the ability to slice the code into cells and to provide additional commentary using Markdown cells. To catch the influence of these notebook features, we developed two new metrics: \textit{Cell coupling} and \textit{Extended comment LOC}.

\textit{Cell coupling}. In our previous work~\cite{titov2021resplit}, we suggested that notebook cells could be perceived as proto-functions, \textit{i.e.}, that they are written to contain a certain logical part of the code and to be reusable. To evaluate this assumption, we developed a coupling metric for cells: we measure how cells are interconnected between each other and compare this value to the traditional coupling of functions. 
More formally, we defined two sets: $C = \{\text{Cells of notebooks}\}$ and $F = \{\text{user-defined functions in a notebook}\}$. So $ \forall c_i \in C $ we define a set of variables used in a cell $c_i$ as $X_i = {\set[x \in c_i]{x \text{ is a variable}}}$ and $\forall f_i \in F $ we define a set of functions that are used within $f_i$ as $Y_i = {\set[y \in f_i]{y \text{ is a function}}}$. Finally, the coupling between functions and the coupling between cells can be calculated as follows:
\shortcouplingformula

\textit{Extended comment LOC}. Notebooks provide users with the ability to write commentaries in a more convenient way via special Markdown cells. This means that the default metric of Comment LOC will not take them into account. To avoid this problem, we developed the \textit{Extended comment LOC} metric, in which we include not only the number of comment lines, but also the number of lines of text in the adjacent Markdown cell. We consider that the Markdown cell could be counted as a comment only to an adjacent cell --- such an approach, adopted from the work of Zhang et al.~\cite{zhang2022coral}, helps us to avoid collecting information unrelated to code. 

The full process resulted in 15 key metrics, their full list can be found in Table~\ref{table:metrics-description}. The table shows the metrics that are implemented at the level of cells and their description. In order to simplify the interpretation of the results, we separated structural metrics into three groups: \textit{code writing} metrics, \textit{function usage} metrics, and \textit{complexity} metrics.

The table also reports the rule of aggregating the metric to the level of an entire notebook. Aggregating is a crucial part for the correct comparison --- most of the metrics were developed to measure structural qualities of the entire code file, while for notebooks, we calculate them for each cell and then aggregate them. Separate calculations for each cell and the following aggregation allow us to gather more data. Some metrics require the code to be syntactically correct in order to be calculated, and analysing the notebook cell-by-cell helps us to calculate the metric for the notebooks that contain cells with incorrect code. Another important step in calculating metric values is their normalization. Because of the possible size difference between notebooks and scripts, we normalized most of the metrics to the number of code lines in the notebook. Therefore, we report most of the code writing metrics and all of the function-based metrics on the \textit{per line} basis.

\subsection{Stylistic Metrics} 
\label{sec:methodology_stylistic}

Structural metrics allow us to evaluate many properties of the code, such as its complexity, the number of various structural elements, and coupling between them. However, these metrics cannot fully tell us how maintainable or error-prone the code is. To answer these questions, software engineers usually employ a set of stylistic metrics that characterize readability and efficiency of code by referring to existing standards (\textit{e.g.}, PEP8~\cite{pep8}). Stylistic errors include, for example, extra indentations in the code, names of functions or variables that are too long, having unreachable parts of the code, useless function calls, etc.

Various libraries implement the calculation of these metrics in different ways, so we decided to use one of the already available tools --- Hyperstyle~\cite{birillo2022hyperstyle}. This tool allows checking the code style using several different style guides: pylint~\cite{pylint}, flake8~\cite{flake8}, and WPS~\cite{wemake}. 
Hyperstyle is able to analyze a separate Python script and return all stylistic errors that are present in it. To process a notebook, we combine all code cells of the notebook in the order of their appearance and pass it as a script to Hyperstyle.

After processing Jupyter notebooks and scripts, we compare them based on the Hyperstyle's own categorization of errors: \textit{Code Style}, \textit{Code Complexity}, \textit{Error-proneness}, \textit{Best Practices}, and \textit{Minor Issues}. We leave out the \textit{Code Complexity} results, because we describe complexity metrics in more detail in the structural part of our study, and \textit{Minor Issues} due to their insignificance. The three remaining categories are focused on code quality: the \textit{Error-proneness} category searches for potential bugs in the code, the \textit{Code Style} category checks if the code is in line with the PEP8 standard, and lastly, the \textit{Best Practices} category checks whether the code is in agreement with the recommended Python practices.

We provide an overview of each category by selecting the most frequent errors using the following procedure:

\begin{enumerate}
    \item For each error, we calculate the frequency of its occurrence in notebooks and scripts separately.
    \item For each error, we calculate the mean of these values.
    \item We sort the errors in a descending order of the mean frequency and take top 5 issues from each category. 
\end{enumerate}

We believe that providing the most frequent errors for both of the studied mediums will help us better understand the specificity of notebooks and the main patterns of their usage. While in the paper, we could not describe the full spectrum of errors, the detailed results are available in our dataset~\cite{dataset}. Importantly, analyzing the results of this comparison allows us to evaluate whether traditional code quality metrics from Python are applicable to notebooks. 

%% file: sections/04-library.tex
\section{\libname}\label{sec:library}

While containing the same Python code, notebooks are structurally different from usual Python modules. This format is relatively new to the Python ecosystem, so very few tools exist for its static analysis~\cite{pimentel2021understanding}. To bridge this gap, conduct our analysis, and provide the research community with a reusable tool, we decided to develop our own library for calculating structural metrics of Jupyter notebooks. In this section, we describe this library in detail.

Storing and analyzing notebooks in their raw form is not very convenient --- their JSON structure contains a lot of data that is unnecessary for the analysis such as ours and consumes a lot of hard drive space. For example, our initial collection of 9 million raw Jupyter files exceeded 4 TB of space. Analyzing such amount of data is a technical challenge that we solved with \libname~\cite{matroskin}.

The \libname library solves the following problems:
\begin{itemize}
    \item Preprocessing Jupyter files into a convenient format.
    \item Calculating the given set of metrics and providing a platform for developing new metrics for Jupyter notebooks.
    \item Storing the preprocessed Jupyter files and their metrics. 
\end{itemize}

\textbf{Preprocessing Jupyter Files.} A Jupyter notebook represents a JSON file with two keys: \textit{metadata} and \textit{cells}. \textit{Metadata} consists of the notebook's kernel specification, such as the programming language, version, file extension, etc. \textit{Cells} contain an array of dictionaries that characterize each of the cells in the notebook. 
For our purposes, we extract two entities from the raw data: (1) the information about the notebooks's programming language, and (2) all the cells in the notebook, \ie the type of each cell and its content. 

\textbf{Calculating the metrics.} Calculating the metrics is a two-step process. The first step is the calculation itself --- \libname contains a runner that takes target metrics as functions, the cell content as input, and calculates the metrics for every eligible cell. As a result, we have raw values of all metrics for the input cells.   
However, the majority of metrics are interpreted on the level of the entire notebook, so we need to aggregate them to perform the analysis (see Section~\ref{sec:methodology_structural}). Thus, the second step is the aggregation. The runner takes the aggregator as a function, the metric and the notebook identifier as input, and returns final values for all the metrics. 

We designed \libname to be as flexible as it can be. The modular structure of the library makes it easy to develop new metrics and aggregation methods. Also, to speed up the analysis, the tool was adapted to work with multiple parallel processes using Ray~\cite{moritz2018ray}. 

Finally, it should be noted that such a calculation process can be easily transferred from notebooks to scripts --- for this, one needs to present the script as a notebook consisting of one cell, thus comparing scripts and notebooks also boils down to comparing uniform data structures.

\textbf{Storing the results.} Lastly, \libname deploys a database to store all the preprocessed notebooks and calculated metrics. The database corresponds to the structure of the library --- it has a table with the notebook data (IDs and aggregated metrics), and a table with the cell data (IDs, content, raw metrics). The preprocessing and the efficient storing of the notebook data resulted in compressing the initial collection of notebooks from 4 TB to just 100 GB.  

%% file: sections/05-dataset.tex
\section{Datasets}\label{sec:datasets}

To carry out the comparison, we collected two different datasets --- Jupyter notebooks and traditional Python scripts. Let us now describe this process in greater detail. 

\subsection{Dataset of Jupyter Notebooks}

\subsubsection{Downloading the data}

The notebooks were collected from GitHub. To collect them, we searched for all notebook files across all public repositories (forks were excluded due to the limitations of GitHub API), which resulted in a list of 1,757,039 projects. We defined notebooks as any files with the \texttt{.ipynb} extension. 
While it is possible for a notebook to have a local import and thus have an extended codebase in the project, we chose not to merge such cases. The basic assumption of our analysis is that writing in a particular environment changes the properties of the code. When someone transfers a part of the code from a notebook into a separate script, we believe that it also implies that some edits will be made, which will change the structure and the style of the code.

The indexing and the downloading of the notebooks took place in September and October of 2020, a total of 9,719,569 notebooks were downloaded. It is important to note that such notebooks can contain code in various languages, not only Python, for example, R, Julia, and others. Since we are interested in Python, we only left the notebooks that had Python declared as a language in their metadata. This resulted in filtering out 13.8\% of the notebooks, leaving 8,381,371 in the initial collection. 

\subsubsection{Extracting the properly licensed dataset}

The majority of the collected notebooks had no open-source license that would cover them. While GitHub regulations~\cite{licenses} allow reporting the aggregated metrics of this data for scientific purposes, other potential applications of the data could be undermined by the absence of a licence.
To create an open dataset that could be used freely in the scientific community, we decided to create a subset of our initial collection that contains notebooks distributed under permissive licenses only. To determine the license of each notebook, we determined the license of the project that it came from. To do that, we used GitHub API that provides the name of the license in the project. Specifically, we selected notebooks from projects with three main permissive licenses: Apache-2.0, MIT, and BSD-3-Clause~\cite{golubev2020study}. These licences allow for a wide range of further applications of the data --- model training, additional research, and source code citation. 
Thus, a license was matched for each notebook, and then the data was split into two subsets: those with suitable licenses and all the rest. In total, the properly licensed dataset included \textbf{847,881} notebooks, \textit{i.e.}, approximately 10\% of all the Python notebooks from GitHub. 

\subsubsection{Analysis of the properly licensed dataset}
After dividing the notebooks into the properly licensed dataset and the rest, we obtained two datasets that are very different in size (847,881 of properly licensed notebooks and 7,533,490 of the remaining ones), so it is important to estimate how much statistically significant information was lost from such a reduction. 

To do that, we performed a statistical comparison of distributions for all metrics described in Section~\ref{sec:methodology} for the two datasets. In the course of this comparison, we found that the mean values and the distributions for the majority of the metrics did not differ between them (p $\leq 0.001 $). We discovered differences in the mean values only for the normalized Comment LOC and the coupling metrics. The average value of Comment LOC in the properly licensed notebooks \APAstats{0.4}{0.34} turned out to be higher than that of the remaining notebooks \APAstats{0.33}{0.32}, \APAt{-99.76}{0.001}. At the same time, there was more coupling in the properly licensed notebooks than in the remaining ones. We believe that these differences are not crucial and do not stop us from using the properly licensed notebooks for further analysis, so from here on out, all the calculations are preformed on the properly licensed dataset, and whenever we refer to the dataset of Jupyter notebooks, we mean the properly licensed notebooks.
To facilitate further research in the area of Jupyter notebooks, we  make this dataset available to the public~\cite{dataset}.

\begin{figure*}[htbp]
  \centering
    \includegraphics[width=\textwidth]{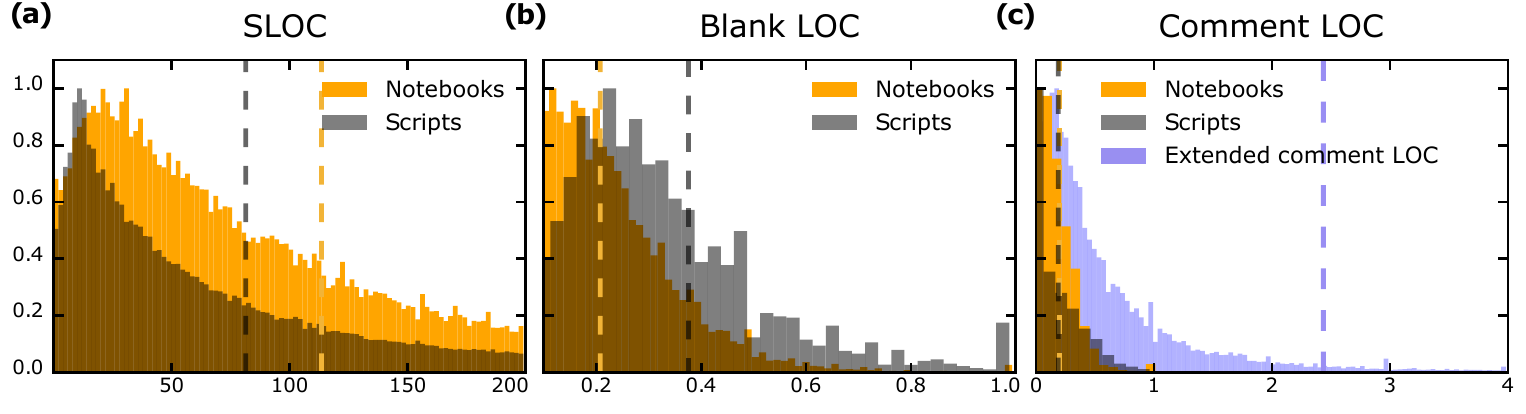}
    \centering
    \caption{Code writing metrics, normalized to their maximum value. Dashed line represents the distribution's mean.}
    \label{fig:code-writing-metrics}
\end{figure*}

\subsection{Dataset of Python Scripts}

To obtain the dataset of Python scripts, we used the following methodology. We used the service developed by Dabic et al.~\cite{Dabic:msr2021data} to obtain the list of all projects that are not forks and have Python as their main language, and then downloaded 10 thousand projects with the most stars. We chose to collect the scripts via filtering by stars because we wanted to acquire more production-like code and avoid as many empty and auxiliary Python scripts as possible.
The main goal of our study is to find the place of Jupyter notebooks as a medium, with the aim of finding possible shortcomings of the current tools and approaches when applied to Jupyter code.
Since a lot of studies use GitHub stars as a proxy metric for the quality of the code~\cite{borges2018s}, we wanted to compare Jupyter notebooks with a dataset like this.
It is worth noting that although such a sample is biased from the standpoint of the project popularity, we assume that such scripts will represent a diverse range of topics since the projects come from completely different domains of software engineering.

This dataset was collected in November of 2021. A special attention was paid to the licenses of the repositories, so we left only the projects with the same three permissive licenses: Apache-2.0, MIT, and BSD-3-Clause. From the 5,133 projects that remained, we extracted all the \texttt{.py} files, excluding the \texttt{\_\_init\_\_.py} and \texttt{setup.py} ones, which resulted in the final dataset of \textbf{465,776} script files.

\subsection{Subsets for the Analysis of Stylistic Metrics}

Lastly, due to the computational complexity of running the Hyperstyle tool, we were constrained in the number of files that we could analyse for stylistic errors. 
We took a subset of 100,000 random files from each dataset --- notebooks and scripts. One restriction for these files was their length --- we took only files with less than $\text{MEAN} + \text{STD}$ lines of code, with MEAN being mean values of code lines in the notebooks and scripts respectively, and STD --- the corresponding standard deviation values.

%% file: sections/06-results.tex
\section{Results}\label{sec:findings}

Let us now describe the results of our study. 
For most metrics values, we report the comparison as the result of a two-sample t-test~\cite{vallat2018pingouin}. We report the mean value and the standard deviation as \textbf{M} and \textbf{SD} respectively for each sample, and t-test as \textbf{t(N)}. 
We chose the significance level $\alpha = 0.005$ and consider the result to be significant only with $p \leq 0.001$.

\subsection{Analysis of Structural Metrics}

Our structural metrics are divided into three groups: code writing metrics, function usage metrics, and complexity metrics. The first group helps us to compare the scale between scripts and notebooks and make sure that the comparison is possible, the second helps us to understand if the profile of function usage differs between the mediums, and the last serves for measuring code complexity in these forms of code representation.   

\begin{figure*}[htbp]
  \centering
  \vspace{0.2cm}
    \includegraphics[width=1\textwidth]{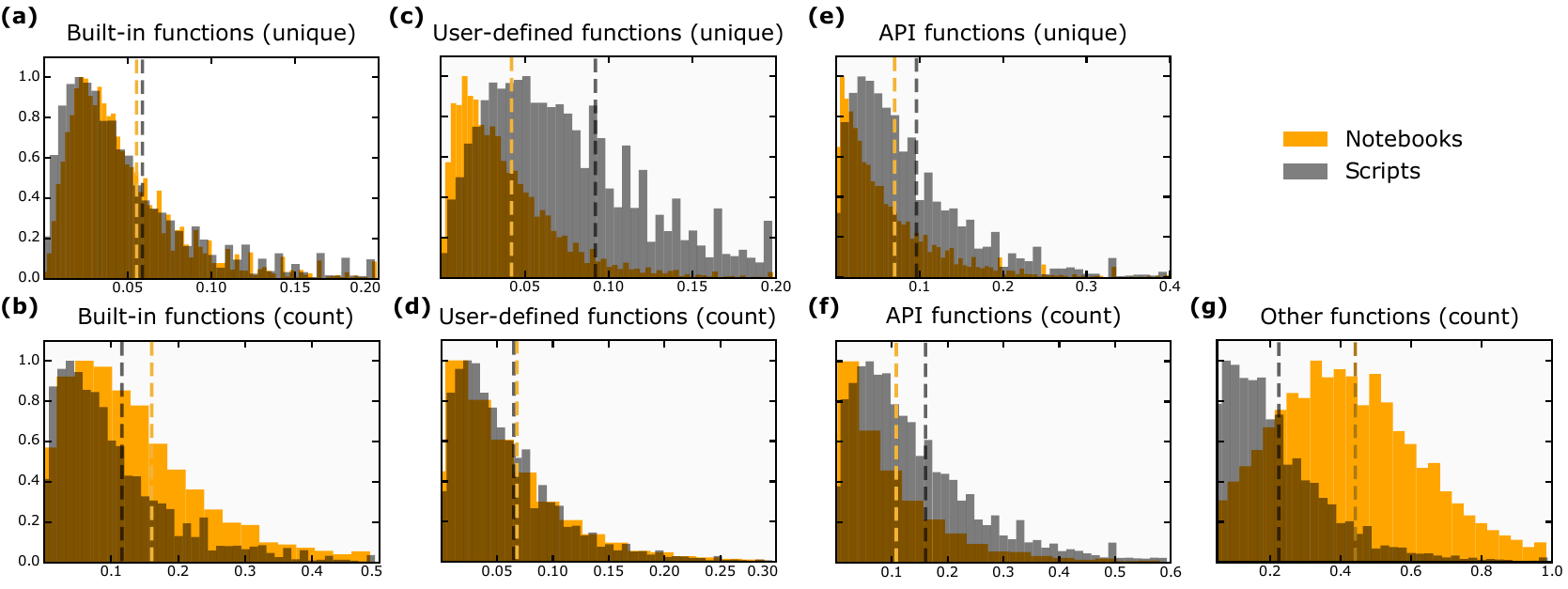}
    \centering
    \caption{Function usage metrics, normalized to their maximum value. Dashed line represents the distribution's mean.}
    \label{fig:functions-using-metrics}
\end{figure*}

\subsubsection{Code Writing}

This group of metrics describes the amount of code in files, as well as its commentary.
Figure~\ref{fig:code-writing-metrics}a shows the distribution of the number of source code lines, \ie the lines of code that are not comments or empty lines. We see that, on average, the notebooks \APAstats{110.0}{160.0} contain 47\% more lines of code than the scripts \APAstats{83.0}{85.0}, \APAt{-143.95}{0.001}. The notebooks are significantly larger than Python scripts and can be extremely long more frequently: 9.1\% of the notebooks contain more than 250 lines of code, while for the scripts this number is only 5.5\%. Since this may cause some metrics to be higher in absolute values, we report further metrics on a \textit{per line} basis when applicable.

Figure~\ref{fig:code-writing-metrics}b shows the distribution of the Blank LOC values and, as we see, the number in the notebooks \APAstats{0.2}{0.31} is less than in the scripts \APAstats{0.37}{1.2}, \APAt{84.26}{0.001}. Such a result could be explained by the fact that notebooks consist of separate cells that split the code into logical parts, while in the scripts, additional blank lines play this role. 

Figure~\ref{fig:code-writing-metrics}c shows the distribution of the Comment LOC values. We see that the number of comment lines in the notebooks \APAstats{0.192}{0.154} is the same as in the scripts \APAstats{0.193}{0.205}, and they do not differ significantly \APAt{2.83}{0.005}. The Extended comment LOC for the notebooks is much higher \APAstats{2.4}{35.0}. This clearly shows that notebooks are more narrative-oriented and contain more information about the code.

\vspace{0.2cm}

\observation{
\textbf{Summary for the code writing metrics.} The notebooks are larger --- they contain on average more code than scripts and more additional information in the form of Markdown cells. This leads us to the conclusion that tools aimed at keeping the code clean and organized are even more crucial for notebooks --- with the increase of the source file size, the probability of an error is also increasing.
}
\subsubsection{Function Usage}

The next group of metrics that we identified characterizes the functions used in the code without taking into account the internal structure of these functions. As we stated in Section~\ref{sec:methodology}, we divided all functions into 4 groups: \textit{Built-in} functions, \textit{User-defined} functions, \textit{API} functions, and \textit{Other} functions. We report the results that relate to these functions normalized to the number of source lines of code in the file.
The distribution of function types is an interesting metric that allows approximating the abstraction level of the code. Chattopadhyay et al.~\cite{chattopadhyay2020s} suggested that notebooks are used mainly for orchestrating API functions and serve as a pipeline for building tools. With our analysis, we provide the validation of this assumption on a larger scale. 

\begin{figure*}[htbp]
  \centering
    \includegraphics[width=\textwidth]{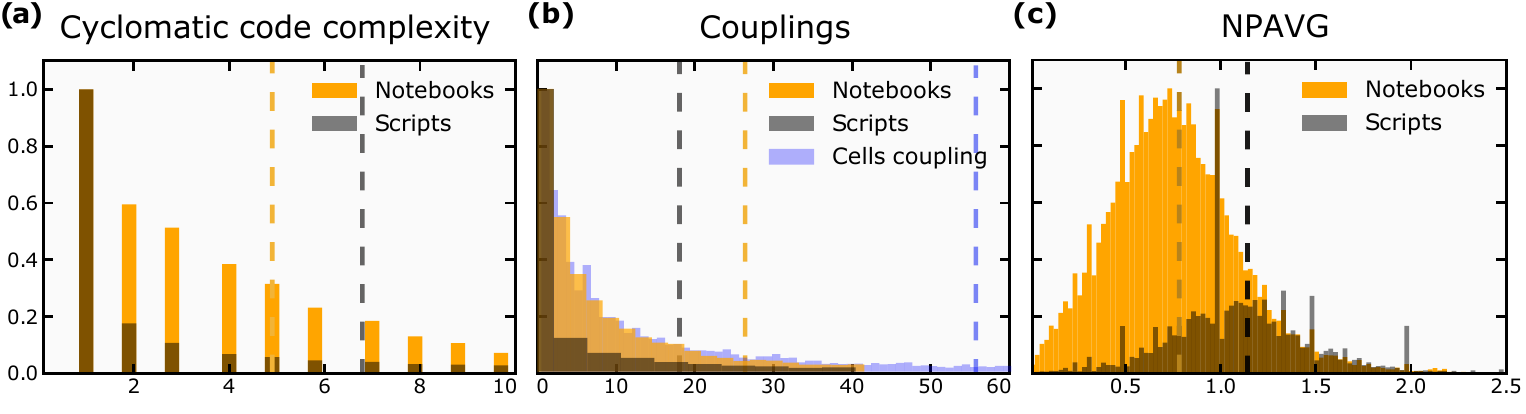}
    \centering
    \caption{Complexity metrics, normalized to their maximum value. Dashed line represents the distribution's mean.}
    \label{fig:complexity-metrics}
\end{figure*}

Figure~\ref{fig:functions-using-metrics}a-b shows the distribution of metrics values for built-in functions. As we can see, although the number of unique built-in functions used is very close in scripts \APAstats{0.057}{0.061} and in notebooks \APAstats{0.055}{0.084}, \APAt{11.64}{0.001}, we observe a significant difference in their total count: it is bigger for the notebooks \APAstats{0.16}{0.26} than for the scripts \APAstats{0.11}{0.099}, \APAt{-142.43}{0.001}. This may be due to the fact that scripts are working with lower level tasks, which require a more diverse set of basic Python constructions, while in notebooks, users tend to repeat the same basic actions, like reading or writing a file. 

Figure~\ref{fig:functions-using-metrics}c-d shows the metrics values for the user-defined functions. We found a significant difference in the number of unique user function definitions: in the notebooks \APAstats{0.042}{0.092} we found much fewer definitions than in the scripts \APAstats{0.085}{0.065}, \APAt{251.09}{0.001}. However, with the total number of usages of these functions, the situation is reversed: the count is bigger for the notebooks \APAstats{0.068}{0.16} than for the scripts \APAstats{0.064}{0.058}, \APAt{-17.22}{0.001}. 
This result lines up with hypotheses suggested in our previous work~\cite{titov2021resplit} that notebook cells can be viewed as proto-functions and serve as the main vessel for the structuring of code, so in notebooks, users tend to define functions only due to the need for a frequent reuse of a particular snippet. 

Figure~\ref{fig:functions-using-metrics}e-f shows the metrics values for API functions. There is a significantly lower number of unique directly imported functions in the notebooks \APAstats{0.069}{0.13} than in the scripts \APAstats{0.098}{0.082}, \APAt{131.53}{0.001}. Respectively, the total count of the usages of such functions is also lower in the notebooks \APAstats{0.11}{0.27} than in the scripts \APAstats{0.16}{0.15}, \APAt{135.30}{0.001}. 
This may be due to the amount of indirect imports in the notebook code. This point could be supported by Figure~\ref{fig:functions-using-metrics}g, which shows the total count for \textit{Other} functions. We found a significant difference between the notebooks \APAstats{0.44}{0.83} and the scripts \APAstats{0.21}{0.16}, \APAt{-249.64}{0.001}: the notebooks have twice as many indirectly imported functions.

\observation{
\textbf{Summary for the function usage metrics.} The patterns of function usage significantly differ between the notebooks and the scripts. The most notable result is the difference between the usage of user-defined functions: notebooks tend to have fewer unique user-defined functions, but use them more frequently. Other results point the way to a lot of potential further research: it seems that notebook users define functions for specific reasons, and the practices of the API methods usage are very distinct for notebooks. These results show that library management and import analysis could be a promising direction for the developers of tools for notebooks.
}

\subsubsection{Complexity}

The last group of structural metrics are complexity metrics. These metrics describe the complexity of the code based on the structural elements used, such as the depth of the code graph, the complexity of functions, as well as the metrics of the relationship between various structural elements. 

Figure~\ref{fig:complexity-metrics}a shows the cyclomatic complexity values of code for the scripts and the notebooks. We can see that the average complexity of the notebooks \APAstats{4.9}{7.1} is less than that for the scripts \APAstats{7.1}{14.0}, \APAt{102.48}{0.001}. This shows us that notebooks tend to have less complex constructions in code. 

Figure~\ref{fig:complexity-metrics}b shows the distribution of coupling values in the scripts and the notebooks. We compare coupling values between functions in the notebooks and the scripts, and additionally we compare these values to the coupling values between cells in the notebooks. These metrics measure the number of shared elements between objects --- functions or cells.  
As we can see, the coupling between functions in the notebooks \APAstats{26.0}{310.0} is 1.5 times greater than in the scripts \APAstats{18.0}{85.0}, \APAt{-14.70}{0.001}. If we consider cells as functions, then the couping between them will be even greater --- \APAstats{53.0}{590.0}. Higher values of coupling could be considered as a sign of entanglement. Ideally, functions should be as independent as possible, but in notebooks, users tend not to separate them that much.

Figure~\ref{fig:complexity-metrics}c shows the distribution of the \npavg. We see that the average value for the notebooks \APAstats{0.78}{0.39} is lower than that for the scripts \APAstats{1.1}{0.47}, \APAt{396.06}{0.001}. Also, we can note a spike at the value of 1. This spike value is the same for both the scripts and the notebooks, which means that functions with a single parameter are significant constructs in both mediums. Some of such scripts represent functionality for visualizing and plotting the data.

\observation{
\textbf{Summary for the complexity metrics.} 
If we combine the results for several metrics, we can say that, on average, functions in the notebooks are more coupled, but they have less complexity and accept fewer parameters as input. We could interpret this as functions in notebooks being simpler but more entangled, with the functionality and fields being less separated. We hypothesize that these findings could be the reason for the observation made by Chattopadhyay et al.~\cite{chattopadhyay2020s} that refactoring is among the most difficult tasks to perform in notebooks. In that case, the works dedicated to finding duplicates or developing the Extract Method refactoring tools are even more important in the context of notebooks.
}
\subsubsection{Summary for the structural metrics}

Structural metrics are difficult to interpret. The results that concern function usage and complexity metrics highlight the specificity of notebook code. Having a lower number of user-defined functions and directly imported ones, we can suppose that notebooks are used more for employing Python modules, importing them fully, and then orchestrating their methods. Despite the fact that the notebook code is less complex, it can be more entangled and thus harder to comprehend. This should be taken into account when developing tools for notebooks --- from the point of the tool's performance and the tool's end goal. Developing tools that could help notebook users to untangle their code seems like a prominent application of our results.

\subsection{Analysis of Stylistic Metrics}

In the next part of the analysis, we measured the code quality using the Hyperstyle tool. As we stated in Section~\ref{sec:methodology_stylistic}, we chose to analyse the following types of stylistic violations: \textit{Error-proneness}, \textit{Code Style}, and \textit{Best Practices}. 

On average, there are \APAstatsNoBraces{184}{244} errors per notebook and \APAstatsNoBraces{126}{355} errors per script. As we noticed before, notebooks tend to be longer than scripts, so we normalized the number of errors to the source lines of code in a notebook or a script, and got the following values: \APAstatsNoBracesMedian{1.06}{2.91}{0.96} for the notebooks, and \APAstatsNoBracesMedian{0.752}{1}{0.533} for the scripts.
Such big values could be explained by the fact that we use the maximum number of available code quality issues to detect, with the total number of unique identifiable issues being more than 600.

\input{sections/05-results-stylystic-table}

In this section, we will analyze the most common errors in more detail, dividing them into the corresponding categories. We report the top five errors for each category with the percentage of files containing each error in Table~\ref{table:stylistic-results}.

\subsubsection{Error-proneness}

In this category, all code quality issues significantly prevail in the notebooks compared to the scripts. We believe that it could be explained by specific features or practices in the notebooks. 
Both of the WPS428 and WPS0104 issues are reporting statements with no effect. Their popularity demonstrates that there is much more code without an effect in the notebooks than in the scripts. This can be explained by the fact that notebooks provide much richer capabilities for the output from code, and many of the statements without an effect could be output functions like \texttt{df.head}. These statements do have an effect, but only for the reader. It could be an important feature for a linter to be able to detect such statements and not report them as code quality issues. 

Both of the WPS442 and WPS440 issues are connected to the reusing of identifiers. In the first case, the bad practice consists of shadowing the identifier defined in the outer scope. Since often notebooks serve as prototyping tools, it is typical for them to use the same placeholder identifier names like \texttt{tmp}, \texttt{test}, or \texttt{x} all over the code. Frequent reuse of identifiers may cause an increase in shadowing. The second issue is more strict, WPS440 (\textit{block variables overlap}) forbids the reuse of identifiers in the current scope. For example, one can catch this issue by using the same identifier \texttt{data} in consequent constructions like \texttt{with open(filename) as data}. 
Lastly, the E0602 issue (\textit{undefined variable}) is also prevailing in notebooks. Here, we can find a simple explanation: notebooks provide the ability to write and execute cells in the arbitrary order, which could result in the non-linear order of code execution. This leads to the linter seeing some variables as undefined, while they just require a specific order of executing cells to be defined. 

\vspace{0.1cm}

\observation{
\textbf{Summary for the \textit{Error-proneness} issues}. In this category, some of the most popular code quality issues are not actually valid issues in the context of notebooks. The WPS428 and WPS0104 issues do not highlight incorrect code behaviour but show rather expected usage of a notebook. However, WPS440 and WPS442 are still demonstrating bad code practices and speak about the lower quality of code in Jupyter notebooks. We conclude that notebooks require specific code quality tools, and these tools should revise some parts of the Python style guide to make it applicable for notebooks.
}

\vspace{0.1cm}

\subsubsection{Code Style}

In this group, most of the issues are associated with an incorrect number of spaces in various places in the code (E266, E231). While these issues do not influence the performance of the code, they might signal the overall lesser quality of Python code in notebooks or the lack of code review. 

The I201 and WPS301 issues are more frequent in the notebooks and are connected to the process of importing modules in Python. One possible explanation for this is that the users import libraries in notebooks on demand and in places where they need a certain library. Writing scripts usually involves more planning and code review, which makes the imports section of the file significantly cleaner and more structured. Additionally, such good results among the scripts may be caused by the fact that modern IDEs are controlling this styling by default, asking the user to correct the issue and even providing an automated fix.   

Lastly, C812 (\textit{missing trailing comma}) is much more frequent in the scripts than in the notebooks. This error is common in a situation when one needs to put data in a Python dictionary --- the trailing comma after the last record in the dictionary is a good practice because it simplifies adding new values in the future. It could be a frequent situation for scripts, where developers keep configurations or other similar data structures. We suppose that people rarely work with such data structures in Jupyter notebooks --- usually all the required data is imported from external files. 

\vspace{0.1cm}

\observation{
\textbf{Summary for the \textit{Code Style} issues}. The results demonstrate that the notebooks show a greater number of code style issues that could be avoided by using simple automated linters, which are usually present in modern IDEs. While some features are available in notebooks, \textit{e.g.}, the automatic conversion from four spaces to a tab, there are a lot of features that need to be incorporated in notebooks, \textit{e.g.}, moving the imports to the first cell of the notebook.
}

\vspace{0.1cm}

\subsubsection{Best Practices}

Two out of the five most popular issues in this category are related to unused imports (W0611, F401). This can be explained by the experimental nature of writing code in Jupyter notebooks. Wan et al.~\cite{wan2019does} suggested that in Data Science tasks, programmers tend to be more cyclical: iterating over and over, trying different solutions, which might cause unnecessary and unused imports.

The frequency of W0621 (\textit{redefining name from outer scope}) could be explained similarly to the WPS440 and WPS442 issues from the Error-proneness category. Reusing the same names is typical for notebooks due to the development speed and the absence of explicit control for such events from the development environment. 

Another error that prevails in the notebooks is related to string concatenation (WPS336). We can guess that such behaviour may be caused by the fact that notebooks provide more ways to output various code results, more code is working directly with strings, and thus will have more issues when manipulating them. 

The R504 error (\textit{unnecessary variable assignment before a return statement}) is the only one among the top five that is more frequent in the scripts. While the difference is not so big, it is significant. The error may be caused simply by a bigger number of unique user-defined functions in the scripts, which could lead to a more frequent situation of unnecessary variable assignment.

\observation{
\textbf{Summary for the \textit{Best Practices} issues}. In the notebooks, we discovered more problems with best practices than in the scripts. Most of the problems highlighted in this part of the analysis arise from the ways notebooks are used, not by the medium itself. Looking at these issues, we can say that it is crucial in the further development of notebook-oriented tools to support the practice of iterative development and create solutions that would help to clean up the code after multiple iterations of changes.%
}

\subsubsection{Summary for the stylistic metrics}
Our analysis showed that Jupyter notebooks contain code of lesser quality: we found significantly more stylistic issues in the notebooks than in the scripts.  However, not all of these issues are actually valid --- some of them are just not applicable for notebooks. For example, WPS428 (\textit{found statement that has no effect}) frequently constitutes outputting something in a notebook. Another batch of issues is very likely caused by the practices of development in notebooks --- iterative development causes some parts of code to become obsolete much faster. 

%% file: sections/05-results-stylystic-table.tex
\begin{table*}[t]

    \begin{tabular}{  c  l  c c }
        \toprule
\textbf{Error}      
& \multicolumn{1}{c}{\textbf{Description}}   
& \textbf{Notebooks (\%)} & \textbf{Scripts (\%)} \\\midrule\midrule
\multicolumn{4}{c}{\textbf{Error-proneness}}\\\midrule\midrule
\textbf{WPS440}  & Found block variables overlap 
& \textbf{41.69} &  19.39 \\\midrule
\textbf{WPS428}  & Found statement that has no effect 
&  \textbf{44.16} &  5.47 \\\midrule
\textbf{WPS0104}  & Statement seems to have no effect
& \textbf{39.8} &  1.11 \\\midrule
\textbf{WPS442}  & Found outer scope names shadowing 
& \textbf{25.15} &  9.12 \\\midrule
\textbf{E0602}  & Undefined variable \%r 
& \textbf{19.44} &  7.37 \\\midrule\midrule

\multicolumn{4}{c}{\textbf{Code style}}\\\midrule\midrule

\textbf{I201}  & Missing newline between sections or imports. 
&  \textbf{75.22} &  40.74 \\\midrule
\textbf{E231}  & Missing whitespace after ',', ';', or ':' 
&  \textbf{59.87} &  11.2 \\\midrule
\textbf{WPS301}  & Found dotted raw import: (imports like ``import os.path``) 
&  \textbf{51.54} &  15.42 \\\midrule
\textbf{E226}  & Missing whitespace around arithmetic operator 
& \textbf{46.51} &  9.55 \\\midrule
\textbf{C812}  & Missing trailing comma 
&  15.86 & \textbf{34.49} \\\midrule\midrule

\multicolumn{4}{c}{\textbf{Best practices}}\\\midrule\midrule
\textbf{F401}  & Module imported but unused 
&  \textbf{47.54} &  14.89 \\\midrule
\textbf{W0611}  & Unused import when preceded by import as 
&  \textbf{47.2} &  13.36 \\\midrule
\textbf{W0621}  & Redefining name from outer scope 
& \textbf{35.72} &  7.05 \\\midrule
\textbf{WPS336}  & Exception for Explicit String Concatenation
&  \textbf{20.48} &  17.74 \\\midrule
\textbf{R504}  & Unnecessary variable assignment before return statement 
& 15.33 &  \textbf{18.07} \\
        \bottomrule
    \end{tabular}
    \vspace{0.3cm}
    \caption{Top five stylistic errors in each category, sorted by the mean frequency of occurrence between both formats. We report the percentage of files with at least one occurrence.}
    \label{table:stylistic-results}
\end{table*}

%% file: sections/07-threats.tex
\section{Threats to Validity}\label{sec:threats}

The general nature of our work leads to certain threats to its validity.
Firstly, this work aims to be an exploratory research, so while we provide statistically significant results, most of our observations should be taken as possible hypotheses and not as conclusions. The main goal of this work is to lay the foundation for further research into the specifics of computational notebooks as a medium. 
    
Another concern is the choice of structural metrics and their adaptation for notebooks. In this work, we implemented a very limited set of metrics in order to process a large amount of data. We hope that in further research, we would be able to take more complex metrics, and the ease of extending \libname can be helpful in this regard. 
Also, this paper is dedicated only to the comparison of metrics between the two codebases, and we did not search for any kind of correlation or interaction between the metrics themselves. While such research was not carried out in this paper, we believe that there could be some interesting interaction effects between some metrics, interactions that are specific to the medium.  
    
There may be some concerns regarding the generalizability of the datasets. The dataset of notebooks could already be considered a little outdated --- completing this research took more than a year after collecting the data, and due to the freshness of the medium, changes of coding practices can happen fast. Also, while we carry out the comparison of the properly licensed dataset with the full collection of notebooks, this sampling still threatens the generalizability of the study. Another concern should be taken into account regarding the dataset of scripts --- we were not able to collect all Python scripts and resorted to collecting only a sample of the Python codebase. This sample could be biased towards the code of higher quality due to the filtration based on the number of GitHub stars. However, the sample is large and covers diverse topics of software engineering.
    
Lastly, our research relies on two complex libraries: \libname and Hyperstyle, that could be vulnerable to a variety of bugs. 
To combat this, we conducted a sanity check for every studied metric and stylistic issue to make sure that they are correctly identified in the notebooks, reviewing a dozen samples per issue. 

Even though these threats are important to note, we believe that they do not invalidate the results of our research.

%% file: sections/08-conclusion.tex
\section{Conclusion \& Future Work}\label{sec:conclusion}

In this work, we carried out a comparison of Python code properties between notebooks and scripts, and found that the notebook code is a little bit simpler, however, more entangled, and usually contains more stylistic errors. 

Firstly, we clearly see a major difference in how user-defined functions are created and used. In notebooks, users less frequently define functions, but when they do, they use them significantly more. Additionally, these functions take fewer arguments than their script counterparts. This demonstrates that the reasons behind the decision to organize the code into functions are different in these mediums.

Secondly, in terms of code quality, we noticed that some of the frequent stylistic issues in notebooks are just the common ways of using notebooks --- like output statements that have no effect or the shadowing of variables. This result is in line with the suggestion made by Pimentel et. al.~\cite{pimentel2021understanding} that notebooks require their own linter, or at least an adapted version of a regular one. Lastly, we hope that providing a properly licensed dataset of Jupyter notebooks could facilitate further research in different domains of software engineering. 

Our study constitutes the first step in researching the specific features of Jupyter notebooks. Several different directions for future work are possible.

\begin{itemize}
    \item Firstly, it is possible to repeat the presented study with even more exhaustive data, overcoming the necessity to limit the number of processed files due to computational constraints, and thus confirming our results.
    \item An entire dimension of comparison that was outside the scope of our work relates to the \textit{collaborative nature} of the studied media. Notebooks and scripts can be compared in terms of the number of commits and authors, the structure of changes in them, etc., and since these media are very different in the way they are utilized, these differences can demonstrate promising results. Some of our results could be explained from this point of view. One may hypothesise that the lower stylistic quality of the notebooks may be caused by a lack of collaboration in the forms of the code review or repository-wide style guides.
    \item Another interesting aspect of comparison is the \textit{domain} where the files come from. In our work, we aim to compare notebooks and scripts in general, whereas selecting scripts solely from the same domains that Jupyter notebooks come from (machine learning, education, etc.) can lead to other interesting insights.
    \item In our work, we hypothesize about certain issues that are present in notebooks. However, it is vital to make sure that these issues are actually issues for people working with the notebooks. For example, while notebooks do contain a lot of duplication and may require significant refactoring, it can be argued that sometimes notebooks are read sequentially and therefore can allow repetition in favour of conciseness. To study this in detail, a survey or an interview with practicing developers is paramount.
    \item On a more technical side, it is of interest to continue the idea of developing notebook-specific linters and smell detectors. A particular area of interest is whether such tools should process and analyze notebooks in the order the cells are written (presented) or in the order they are executed.
\end{itemize}

With this work, we wanted to lay a foundation for further research and development of computational notebooks. We hope that the conclusions that we made in this paper will highlight the unique properties of computational notebooks as a medium for code, and will help with the development of notebook-specific tools and further research of this medium.

\section*{Acknowledgements}

We would like to thank Anastasiia Birillo for helping us with running the Hyperstyle tool on Jupyter notebooks, our colleagues from the development team of Datalore for helping with collecting the dataset of Jupyter notebooks, and to anonymous reviewers for providing insightful ideas for improving the paper.